\documentclass[12pt]{article}

\begin{document}
\setcounter{page}{1}

\baselineskip=15.5pt \pagestyle{plain} \setcounter{page}{1}
\begin{titlepage}
\begin{center}
\vspace*{-3cm}  \hfill UFR-HEP/07/98\\ {}~{} \hfill \\ \vskip 2cm
{\LARGE {\bf Fractional Supersymmetry\\
                        As\\\vskip 0.5cm A Matrix Model
}} \vskip 1 cm {\large Ilham Benkaddour and El Hassane Saidi }
\vskip .8cm
   UFR-High Energy Physics, Department of Physics, Faculty of Sciences,\\
    Rabat University, Av. Ibn Battouta B.P. 1014, Morocco.
\vskip 0.5cm {\tt H-saidi@fsr.ac.ma }\\
 \today
\end{center}
\vspace{5mm} {\bf Abstract}\\
 Using parafermionic field
theoretical methods, the fundamentals of 2d fractional
supersymmetry ${\bf Q}^{K} =P$ are set up. Known difficulties
induced by methods based on the $U_{q}(sl(2))$ quantum group
representations and non commutative geometry are overpassed in the
parafermionic approach. Moreover we find that fractional
supersymmetric algebras are naturally realized as matrix models.
The K=3 case is studied in details. Links between 2d $({1\over
3},0)$ and $(({1\over 3}^{2}),0)$ fractional supersymmetries and
N=2 U(1) and N=4 su(2) standard supersymmetries respectively are
exhibited. Field theoretical models describing the self couplings
of the matter multiplets $(0^{2},({1\over 3})^{2},({2\over
3})^{2})$ and $(0^{4},({1\over 3})^{4},({2\over 3})^{4})$ are
given.\\ \\ PACS numbers: 0220, 1110K, 1125H, 1130P\\
\\ Published in Class.Quantum Grav. {\bf
16}(1999)1793-1804
\end{titlepage}

\newpage 

\section{Introduction}

In the few last years, there has been attempts to develop a
superspace formulation of 2d QFT that is invariant under
fractional supersymmetry (FSS) defined by:$Q^{k}=P,\quad
\bar{Q^{k}}=\bar{P},\quad K>2$, together with relations involving
both Q and $\bar{Q}$ . In these Eqs $P=P_{-1}$ and
$\bar{P}=P_{+1}$ are the two heterotic components of the 2d energy
momentum vector operator $P_{m}$;\quad $Q\equiv Q_{-1/k}$ and
$\bar{Q}\equiv Q_{1/k}$ are the basic generator of FSS which carry
a 2d spin $s=1/k$ . Most of the studies of this special class of
2d massive QFT are essentially based on methods of  $U_{q}(sl(2))$
quantum group representations and non commutative geometry [1,2] .
This approach however met many difficulties and has lead to
partial results only. Among these difficulties we quote the two
following ones associated with the heterotic K=3 case $(Q^{3}=P)$
. The first one deals with the computation of the two points
correlation function $<\Psi _{-1/3}(z_{1}),\Psi _{-2/3}(z_{2})>$
of the two partners $\Psi _{-1/3}(z)$ and $\Psi _{-2/3}(z)$ of the
bosonic field $\varphi(z)$ of the scalar representation $(\varphi,
\Psi _{-1/3},\Psi _{-2/3})$  of the $(Q^{3}=P)$ algebra. From the
view point of the 2d conformal field theory, one expects from the
braiding feature $z_{12}\Psi _{-1/3}(z_{1})\Psi
_{-2/3}(z_{2})=z_{21}\Psi _{-2/3}(z_{2})\Psi _{-1/3}(z_{1})$, that
the field operators  $\Psi _{-1/3}$ and $\Psi _{-2/3}$ should
anticommute. This result however is not fulfiled in the approach
based  on the $U_{q}(sl(2))$ quantum groups and non commutative
geometry where the $\Psi _{-1/3}$ and $\Psi _{-2/3}$ fields obey a
generalised commutation rule [3] leading to a two point
correlation function $<\Psi _{-1/3}(z_{1}),\Psi
_{-2/3}(z_{2})>_{U_{q}(sl(2))}$ which  violate locality.\\
 The second thing we want to quote concerns the construction of the
generalised superspace and superfields. In this context, one uses
generalised Grassmann variables $\theta_{1}= \theta_{1/3}$ and
$\theta_{2}= \theta_{-1/3}$ satisfying a third order nilpotency
condition $\theta_{\pm 1/3}^{3}=0$, together with  $\theta_{\pm
1/3}^{2}\neq 0$ and $\theta_{1}\theta_{2}=\omega
\theta_{2}\theta_{1}$, where $\omega$ is a C-number such that
$\omega ^{3}=1$. The condition $\theta_{\pm 1/3}^{3}=0$, which
generalizes the usual condition of Grassmann variables  of spin
1/2 , is necessary in order to describe off shell representations
of the fractional supersymmetric algebra (FSA) in terms of
superfields $\Phi(z,\theta ,\bar{z},\bar{\theta})$. In this
language, FSS is generated by translations along the $\theta $ and
$\bar{\theta}$ directions, i.e. $\theta \rightarrow \theta
+\varepsilon$ and  $\bar{\theta} \rightarrow \bar{\theta}
+\bar{\varepsilon}$, where $\varepsilon$ and $\bar{\varepsilon}$
are generalized Grassmann variables of same nature as $\theta $
and  $\bar{\theta}$ . Until now things are quite similar to the
superspace formulation of ordinary supersymmetry associated with
K=2. However there are weak points in the construction of the
generalised superspaces for $K>2$. One of these weakness  deals
with the treatemenent of $\theta ^{2} $,  $\bar{\theta}^{2}$ and
the  $\varepsilon$ and $\bar{\varepsilon}$ parameters. In trying
to establish a given relation by following two equivalent paths,
we get different results. On one hand, $\theta (\bar{\theta})$
should commute with itself as required by the identity
$(1-x)\theta  ^{2}=0$ , which implies that either $\theta ^{2}=0$
and $x\neq 0$ or  $x= 1$ and $\theta ^{2}\neq 0 $. On the other
hand $\theta (\bar{\theta})$ and $\varepsilon(\bar{\varepsilon})$
are required to satisfy generalized commutation rules type $\theta
\varepsilon=\omega \varepsilon \theta(\bar{\theta}
\bar{\varepsilon}=\omega \bar{\varepsilon} \bar{\theta})$; $\omega
^{3}=1$. This is a contradiction since $\theta (\theta +
\varepsilon)\neq \omega (\theta + \varepsilon) \theta$. This
difficulty may be also viewed from the generalised commutation
rule $\theta \eta =\omega \eta \theta$. Taking the limit $\eta =
\theta$, one obtains $\theta ^{2}=\omega \theta ^{2}$  which
implies that  $\theta ^{2}$ should be zero.\\ The aim of this
paper is to set up the basis of 2d fractional supersymmetric QFT
by using parafermionic field theoretical methods. These methods,
which were considered recently in [4], seems to be the right
language to develop a local fractional supersymmetric QFT. This
believe is also supported by the fact that after all fractional
supersymmetry  is nothing but a residual symmetry of massive
perturbations of parafermionic critical models. From this point of
view; the generators of fractional supersymmetry are just remanant
constants of motion surviving after deformations of parafermionic
conformal invariance. Thinking of 2d fractional supersymmetry as a
finite dimensional subsymmetry of the $Z_{K}$ conformal invariance
[5] for instance, one discovers that all the known difficulties of
the abovementioned approach disappear. Moreover we find that
fractional supersymmetric algebras are generated by more than one
charge operator and can be described in a natural way as a matrix
model. The charge operators $Q_{-x},\quad x=0, 1/k, \ldots$, form
altogether a $K\times K$ matrix operator  allowing to define the
fractional supersymmetric algebra as:\\
\begin{equation}
trQ^{k}=P_{-1},
\end{equation}
together with other relations. For details see Eqs (5); (8) and
(17).\\This paper is organized as follows. In section 2, we study
the matrix realisations of 2d supersymmetry. In section 3, we work
out the links between fractional supersymmetry and parafermions.
Matrix realisations of fractional supersymmetry are analysed in
sections 4 and 5. In section 6, we describe brefly the superfield
theory of the matter couplings of $({1\over 3},{1\over 3})$ and
$(({1\over 3})^{2},({1\over 3})^{2})$ fractional superalgebras. In
section 7, we present discussions and give our conclusion.

\subsection{2d supersymmetry as a matrix model}
In this paragraph we consider the 2d heterotic  supersymmetric
algebras respectively defined  as:
\begin{equation}
\begin{array}{lcr}
\{Q_{-1/ 2},Q_{-1/ 2}\}&=& P_{-1}\nonumber \\ \lbrack
P_{-1},Q_{-1/ 2}\rbrack &=&0,
\end{array}
\end{equation}
and
\begin{equation}
\begin{array}{lcr}
\{Q_{-1/ 2}^{+},Q_{-1/ 2}^{-}\}&=& P_{-1}\nonumber \\ \{Q_{-1/
2}^{\pm},Q_{-1/ 2}^{\pm}\}&=& 0\nonumber \\
 \lbrack P_{-1},Q_{-1/2}^{\pm}\rbrack &=&0,
\end{array}
\end{equation}
 and shows that they can be represented as $2\times 2$ matrix models.
A generalisation to higher dimensional spaces of these
representations turns out to offer a natural framework to study
fractional supersymmetry. It shows moreover that methods of
standard supersymmetry can be also used to deal with exotic
supersymmetries.\\ \quad To start consider the following off
diagonal symmetric $2\times 2$ matrix {\bf Q} whose  entries are
given by the 2d $({1\over 2},0)$ supersymmetric generator $Q_{-1/
2}$:
\begin{equation}
{\bf Q}=\left[
\begin{array}{ccccc}
0 & Q_{-1/ 2}\\ Q_{-1/ 2} &0
\end{array}
\right]  \label{matrices}
\end{equation}
This matrix operator acts on the 2d space of quantum states
$|B\rangle =B(0)|0\rangle $ and $|F\rangle =F(0)|0\rangle $ where
the letters B and F stand for Bose and Fermi fields.Taking the
square product of Eq (4), it is not difficult to see that the
$({1\over 2},0)$ supersymmetric algebra (2) may be defined as:
\begin{equation}
\begin{array}{lcr}
tr{\lbrack {\bf Q}^{2}\rbrack }&=& P_{-1}\nonumber \\ tr{\lbrack
{\bf Q} \rbrack }& =&0,
\end{array}
\end{equation}
where the suffix tr means the usual matrix trace operation. There
are several remarkable features of the matrix  definition Eqs
(4-5) of  the 2d $({1\over 2},0)$ supersymmetric algebra; some of
them are manifest at the level of  the $K\times K=2\times 2$
matrix model, others are hidden and emerge for $K\geq 3$
representations.  For details see sections 3 and 4.\\ Moreover
combining the usual constraints of  2d supersymmetry, in
particular hermiticity of the energy momentum vector $P_{-1},
P_{-1}^{+}=P_{-1}$ with the features of the matrix model (5), one
sees that the realisation of {\bf Q}, Eq (4), is not the unique
one. Indeed, decomposing {\bf Q} as ${\bf Q}={\bf Q}^{+}+{\bf
Q}^{-}$
\begin{equation}
{\bf Q}^{+}=\left[
\begin{array}{ccccc}
0 & Q_{-x}^{+}\\ 0 &0
\end{array}
\right] ;\quad \quad {\bf Q}^{-}=\left[
\begin{array}{ccccc}
0  & 0 &  \\Q_{-y}^{- }  & 0,
\end{array}
\right]  \label{matrices}
\end{equation}
one sees that the spins x and y carried by the $Q^{\pm}$ charge
operators are related as
\begin{equation}
x+y=1,
\end{equation}
as required by the first relation of Eq (5). Abstraction done from
Eq (2) and (3), there are infinitely  many solutions of Eqs (6
-7); we shall first consider the two solutions $x=y={1\over 2}$
and x=0 and y=+1 related with the leading zero mode operators of
the NS and Ramond sectors of 2d superconformal invariance. Later
on we shall explore the interesting cases $x={1\over k}$ and
$y={(k-1)\over k}, k=3,4,\ldots ,$ and their link with $Z_{K}$
parafermionic invariance.Now using  Eqs (6) instead of Eq (4) one
finds that the algebra Eqs (5) reads in general as:
\begin{equation}
\begin{array}{lcr}
tr{\lbrack {\bf Q}^{+}{\bf Q}^{-}\rbrack }= P_{-1}\nonumber \\
tr{\lbrack {\bf Q}^{+}{\bf Q}^{+}\rbrack }=tr{\lbrack {\bf
Q}^{-}{\bf Q}^{-}\rbrack }&=& 0\nonumber \\ tr{\lbrack {\bf
Q}^{+}\rbrack }=tr{\lbrack {\bf Q}^{-}\rbrack } &=&0.
\end{array}
\end{equation}
Eqs (8) has a U(1) automorphism symmetry which breaks down to
$Z_{2}$ in the case of Eqs (5) , this is why we shall refer to Eqs
(5) and Eqs (8) as 2d $({1\over 2},0)$ and $(({1\over 2})^{2},0)$
supersymmetric algebras respectively. Moreover introducing the
generators of the gl(2) Lie algebra, we have the following
$2\times 2$ matrix representations:\\
 (i). the NS like superalgebra.\\
In this case the supersymmetric $2\times 2$ matrix generators
$Q^{\pm}$ read as:
\begin{equation}
\begin{array}{lcr}
 {\bf Q}^{+}&=& Q_{-1/2}E_{12}\nonumber, \\
 {\bf Q}^{-}&=& Q_{-1/2}E_{21} ,
\end{array}
\end{equation}
for $({1\over 2},0)$  supesymmetry and
\begin{equation}
\begin{array}{lcr}
 {\bf Q}^{+}&=& Q_{-1/2}^{+}E_{12}\nonumber, \\
 {\bf Q}^{-}&=& Q_{-1/2}^{-}E_{21} ,
\end{array}
\end{equation}
for $(({1\over 2})^{2},0)$  supersymmetry.\\ (ii). The Ramond like
algebra.\\ Here the matrix generators are realized as:
\begin{equation}
\begin{array}{lcr}
 {\bf Q}^{+}&=& Q_{0}E_{12}\nonumber, \\
 {\bf Q}^{-}&=& Q_{-1}E_{21} ,
\end{array}
\end{equation}
for the hermitian case
\begin{equation}
\begin{array}{lcr}
 {\bf Q}^{+}&=& Q_{0}^{+}E_{12}\nonumber, \\
 {\bf Q}^{-}&=& Q_{-1}^{-}E_{21} ,
\end{array}
\end{equation}
for the complex one. Note that compatibility between the usual
hermiticity (+) of the energy momentum vector $P_{-1 }$ and the
adjoint conjugation ({\bf +}) of the $Q^{\pm}$ matrix generators
requires the following identifications:
\begin{equation}
\begin{array}{lcr}
{\lbrack  Q_{-x}^{\pm}\rbrack }^{\bf +}= Q_{-1+x}^{\mp}\nonumber
\\{\lbrack  Q_{-x}\rbrack }^{{\bf +}}= Q_{-1+x}
\end{array}
\end{equation}
Eqs (11) show that $Q_{-1/2}$ is self adjoint whereas $Q_{0}$ and
$Q_{-1}$ are interchanged under the ({\bf +}) conjugation. However
knowing that $Q_{0}$ should satisfy a 2d Clifford algebra
$\{Q_{0},Q_{0}\}=1$, and using the current modes of the Ramond
superconformal algebra [6], one may set
$2Q_{-1=}\{Q_{0},P_{-1}\}=2Q_{0}P_{-1}$ . This identification
reduces the number of generators of the algebra Eq (11). Likewise,
one may also set $Q_{-1}^{-}=Q_{0}^{-}P_{-1}$ where $Q_{0}^{-}$
together with $Q_{0}^{+}$ satisfy
\begin{equation}
\begin{array}{lcr}
\{Q_{0}^{+},Q_{0}^{-}\}&=&1\nonumber
\\\{Q^{\pm},Q^{\pm}\}&=&0,
\end{array}
\end{equation}
and where $Q_{0}=Q_{0}^{+}+Q_{0}^{-}$. Before going ahead note
that the 2d NS like superalgebra generated by
$Q_{-1/2}(Q_{-1/2}^{\pm})$ exchanges bosons B and fermions F
whereas the Ramond like one generated by $Q_{0}(Q_{0}^{\pm})$
preserves the statistics. In the 2d quantum field language, the
Ramond like algebra acts on the field doublet $(\Phi
_{1/2}(z),\Phi _{-1/2}(z)),\quad Q_{0}^{\pm},\Phi _{\pm
1/2}=0;\quad Q_{0}^{\pm}\Phi _{\mp 1/2}=\Phi _{\pm 1/2}$ , as
follows:
\begin{equation}
{\Phi _{\pm 1/2}}{\longrightarrow _{Q_{0}^{\mp} }}\quad {\Phi
_{\mp 1/2}}{\longrightarrow _{Q_{0}^{\mp} }}\quad
\partial _{z}\Phi _{\pm 1/2}.
\end{equation}
 Note, moreover  that the NS and R like
algebras can be realized by help of Grassman variables $\theta
_{\pm 1/2}^{\pm}$ and $\theta _{0}^{\pm}$ as:
\begin{equation}
\begin{array}{lcr}
Q_{-1/2}^{\pm}&=&{\partial \over {\partial \theta
_{1/2}^{\mp}}}+{1\over 2}\theta _{1/2}^{\pm}{\partial _{-1}}
\nonumber
\\Q_{0}^{\pm}&=&{\partial \over {\partial \theta
_{0}^{\mp}}}+{1\over 2}\theta _{0}^{\pm}.
\end{array}
\end{equation}
Similar realizations may be also written down for the hermitian
charge operators $Q_{-1/2}$ and $Q_{-1/2}$. In what follows, we
want to extend the above matrix formulation to the case K=3. At
first sight this should be possible if one succeeds to relate
fractional supersymmetry to an infinite dimensional fractional
superconformal invariance in the same manner as usual  2d
supersymmetry is related to the superconformal invariances. To
that purpose  we start by establishing the link between fractional
supersymmetry and the $Z_{K}$ parafermionic invariance. Then we
study its matrix realisation.
\section{ Fractional supersymmetry and parafermions}
 Here we want to show that from the point of view
of 2d $Z_{K}$ parafermionic invariance, the definition of
fractional supersymmetry, say $(Q_{-{1/k}})^k=P, K>2$, is just a
formal one. The right way to define it is as in Eqs (1). The
latter show
 that FSS is in fact generated by many charge operators which we denote as
 $Q_{-x}, x=0,1/k,\ldots $.More precisely FSS is generated by a
 $k\times k$ matrix ${\bf Q}$ whose entries are the charge operators
 $Q_{-x}$ carrying various values of fractional spins. For the
 $k=3$ case we are interested in here, there are in total three
 pairs of charge operators $Q_{-x}^{\pm},x=0,1/3,2/3$, together
 with $P_{-1}$. Altogether, these operators generate the formal
 cubic nonlinear algebra $Q^{3}=P_{-1}$ which as we shall see, is
 correctly defined as
\begin{equation}
\begin{array}{lcr}
tr{\lbrack  {\bf Q}^{+}{\bf Q}^{-}\rbrack }&=&P_{-1}\nonumber
\\tr{\lbrack  {{\bf Q}^{+}}^{2}\rbrack }=tr{\lbrack  {{\bf Q}^{-}}^{2}\rbrack }
&=&0\nonumber\\tr{\lbrack  {{\bf Q}^{+}}\rbrack }=tr{\lbrack {{\bf
Q}^{-}}\rbrack } &=&0.
\end{array}
\end{equation}
In equation (12) $Q^{\pm}$ are $3\times 3$ matrices given by:
\begin{equation}
{\bf Q}^{+}=\left[
\begin{array}{ccccc}
0 & 0 & Q_{-1/3}^{-}\\Q_{-2/3}^{-} & 0 &0\\ 0 & Q_{-1}^{-} & 0\\
\end{array}
\right] ;\quad \quad {\bf Q}^{-}=\left[
\begin{array}{ccccc}
0 & Q_{-1/3}^{+} & 0 &  \\0 & 0 &Q_{0}^{+ }\\Q_{-2/3}^{+} & 0 &0 ,
\end{array}
\right]  \label{matrices}
\end{equation}
Note by the way that here also $Q_{-x}^{+}$ and $Q_{-y}^{-}$ obey
Eqs (13) and $Q_{-1}^{-} = Q_{0}^{-} P_{-1}$. A way to derive Eqs
(17-18) is to consider the $Z_{3}$ parafermionic invariance of
Zamolodchikov and Fateev (ZF) [5] given by :
\begin{equation}
\begin{array}{lcr}
\Psi ^{\pm}(z_{1})\Psi ^{\pm}(z_{2})\approx  -z_{12}^{-2/3}\Psi
^{\pm}(z_{2})\nonumber
\\\Psi ^{+}(z_{1})\Psi ^{-}(z_{2})\approx  -z_{12}^{-4/3}{\lbrack 1+5/3
z_{12}^{2}T(z_{2})\rbrack }\nonumber
\\T(z_{1})\Psi ^{\pm}(z_{2})\approx  {2/3\over z_{12}^{2}}\Psi ^{\pm}(z_{2})
+{1\over z_{12}}\partial _{z}\Psi ^{\pm}(z_{2})\\
T(z_{1})T(z_{2})={k-1\over
{k+2}}z_{12}^{-4}+2z_{12}^{-2}T(z_{2})+z_{12}^{-1}T(z_{2}).\nonumber
\\
\end{array}
\end{equation}
In these Eqs $T(z)$ is the usual energy momentum current, $\Psi
^{\pm}(z)([\Psi ^{\pm}(z)]^{+}=\Psi ^{\mp}(z))$ are the ZF
parafermionic currents of spin $2\over $. The algebra (19) has
three parafermionic highest weight representations (PHWR)$[\Phi
_{q}^{q}]; q=0,1,2.$ namely the identity family $I=[\Phi
_{0}^{0}]$ of highest weight $h_{0}=0$ and two degenerate families
$[\Phi _{1}^{1}]$ and $[\Phi _{2}^{2}]$ of weights
$h_{1}=h_{2}={1\over 15}$.
 Each one of these PHWR $[\Phi _{q}^{q}]$ is reducible into three Virasoro
 HWR $(\Phi _{q}^{p}), p=q, p=q+mod6$. For more details on the representation
 theory of Eqs  (18)  see [5-7]. All we need in the achievement of this
 study is  the algebra Eqs (19),the identities
\begin{equation}
\begin{array}{lcr}
\Psi ^{\pm} \times \Phi _{q}^{p}=\Phi _{q}^{p\pm 2}\quad (a)\nonumber \\
 \Phi _{q}^{p\pm 6}=\Phi _{q}^{p}\quad \quad \quad \quad (b)\\,
\end{array}
\end{equation}
as well as the following braiding property of the conformal field
operators $\Phi _{1}(z_{1})$ and $\Phi _{2}(z_{2})$:
\begin{equation}
z_{12}^{\Delta}\Phi _{1}(z_{1})\Phi _{2}(z_{2})=z_{21}^{\Delta}
\Phi _{2}(z_{2})\Phi _{1}(z_{1})=\Phi(z),
\end{equation}
where $\Delta =\Delta _{1}+\Delta _{2}+\Delta _{3},\Delta _{i},
i=1,2,3$, are the weights of the $\Phi _{i}$'s. Note that Eq (20-a)
is a formal Eq which should be understood as:
\begin{equation}
\Psi ^{\pm}(z_{1})\Phi _{q}^{p}=\sum z_{12}^{n-1\mp p/3}Q_{-n\pm (p\pm 1)/3}^
{\pm} \Phi _{q}^{p}(z_{2}),
\end{equation}
where $Q_{-n\pm (p\pm 1)/3}^{\pm}$ are the mode operators of $\Psi ^{\pm}$
which in turn may be defined as
\begin{equation}
Q_{-n\pm (p\pm 1)/3}^{\pm}|\Phi _{q}^{p}\rangle =\oint dzz^{n\pm p/3}
 \Psi ^{\pm}(z)\Phi _{q}^{p}(0)|0\rangle .
 \end{equation}
In addition to Eqs (20) which predicts the existence of three
doublets of the charge operators $Q_{-x}^{\pm},x=0,1/3,2/3$, one
also has the braiding feature (21) playing a crucial role  in the
building of FSS. For example the deformation parameter $\omega $
(often denoted by q in the literature ) of quantum groups and non
commutative geometry dealing with fractional supersymmetry should
be related with the braiding property of the conformal field
blocks $\Phi _{1}(z_{1})$ and $\Phi _{2}(z_{2})$. Eq (21) tells us
that the parameter $\omega $ is equal to $exp(\pm i\pi \Delta)$.
For $\Delta _{3}=0$ for instance, $\omega $ reduces to $exp(\pm
i[\pi (\Delta _{1}+\Delta _{2})])$, so that for $\Delta
_{1}+\Delta _{2}=1$, the field operators $\Phi _{1}$ and $\Phi
_{2}$ anticommute and then should be treated as fermions. Choosing
$\Delta _{1}=1/3$ and $\Delta _{2}=2/3$ for example, the two point
function $<\Phi _{1/3}(z_{1},\Phi _{2/3}(z_{2}> $ should be equal
to $-<\Phi _{2/3}(z_{2},\Phi _{1/3}(z_{1}>=1/z_{12} $. This
result, derived from parafermionic conformal field methods, solves
the difficulty of refs [3] according to which the two point
function $<\Phi _{1/3}(z_{1},\Phi _{2/3}(z_{2}> _{U_{q}(sl(2)}$
computed in a model of fractional supersymmetry based on quantum
groups and non commutative geometry methods; i.e:
\begin{equation}
<\Phi _{1/3}(z_{1},\Phi _{2/3}(z_{2}> _{U_{q}(sl(2)}=e^{2i\pi /3}
<\Phi _{2/3}(z_{2},\Phi _{1/3}(z_{1}> _{U_{q}(sl(2)}={1\over
z_{12}}.
\end{equation}
Eq (24) shows that the model based on the $U_{q}(sl(2)$ methods is
non local. In summary local 2d field theoretical realisation  of
fractional supersymmetry cannot be generated by only one charge
operator. The number of generators may be obtained by analysing
the mode operators $Q_{-n\pm (p\pm 1)/3}^{\pm}$, n integer. The
$Q_{-n\pm (p\pm 1)/3}^{\pm}$'s depend on the p charge of the
conformal representation $|\Phi _{q}^{p}\rangle $  on which they
act. For q=0 for example, the non vanishing actions of
$Q_{-x}^{\pm},x=0,1/3,2/3$ on the leading states $|s,p\rangle $ of
spin s, $0\leq s\leq 1$  and charge p read as:
\begin{equation}
\begin{array}{lcr}
Q_{-2/3}^{\pm}|0,0\rangle &=&|{2\over 3},0\rangle \nonumber \\
Q_{0}^{+}|{2\over 3},+2\rangle &=&|{2\over 3},-2\rangle \nonumber
\\ Q_{0}^{-}|{2\over 3},-2\rangle &=&|{2\over 3},+2\rangle
\nonumber\\ Q_{-1/3}^{+}|{2\over 3},-2\rangle &=&|1,0\rangle\\
Q_{-1/3}^{-}|{2\over 3},+2\rangle &=&|1,0\rangle .\nonumber \\
\end{array}
\end{equation}
From these Eqs and Eqs (22-23), one sees that $Q_{-1/3}^{\pm}$ and
$Q_{0}^{\pm}$ cannot act directly on the state $|0,0\rangle $
similarly $Q_{-2/3}^{\pm}$ cannot operate directly on $|{2\over
3},\pm 2\rangle $. This feature gives another indication that FSS
should be generated by more than one Q operator as it is currently
used in the literature based on quantum groups and non commutative
geometry approaches. Recall that the first indication we have
mentionned in the introduction of this paper refers to the
inconsistences induced by the introduction of the generalized
Grassmann variable $\theta $ satisfying a higher order nilpotency
condition $\theta ^{k}=0, k>2$ with $\theta ^{k-1}\neq 0 $ and
where the problem of locality rised above is just one of the
manifestation of the limit of the methods used. In the approach
based on parafermionic conformal field theorical techniques we are
considering here, these problems are overpassed. Locality is
restored since all fields carrying fractional spins obey an
anticommuting statistics and the higher order Grassmann nilpotency
necessary for the description of off shell representations of FSS
is ensured by the presence of more than one charge operator.
Having discussed the link between FSS and parafermions, we turn
now to study its relation with matrix theory.
\section{Fractional supersymmetry as a matrix model}
 Starting from Eqs (25) and denoting by $\Pi _{r} ,r=0,\pm 1$, the projectors
along the states $|s,p\rangle =|i\rangle (p=2i)$ and by
$E_{ij}=|i\rangle {\otimes } |j\rangle $  the generators of the
gl(3) Lie algebra rotating the state $|i\rangle $ into the state
$|j\rangle , i,j=0,\pm 1$, (the indices $\pm $ refer to the two
fractional supersymmetric partners of the bosonic state indexed by
i=0.), one sees that the component $P_{-1}$ of the energy momentum
vector reads in terms of the $Q_{-x}^{\pm}$'s,$x=0,1/3,2/3$ and
the projectors as:
\begin{equation}
\begin{array}{lcr}
P_{-1}\approx tr{\bf P} \nonumber \\ {\bf
P}=Q_{-1/3}^{+}Q_{0}^{+}Q_{-2/3}^{+}\Pi _{0}+
Q_{-2/3}^{+}Q_{-1/3}^{+}Q_{0}^{+}\Pi
_{1}+Q_{0}^{+}Q_{-2/3}^{+}Q_{-1/3}^{+}\Pi _{-1}.
\end{array}
\end{equation}
Note that  a similar relation to this Eq using $Q_{-x}^{-}$
instead of $Q_{-x}^{+}$ is also valid . Moreover using  Eqs (25),
one many rewrite Eqs (26) and its hermetic conjugate in the
following form.
\begin{equation}
2P =\lbrack
Q_{-1/3}^{+}Q_{-2/3}^{-}+Q_{-1/3}^{-}Q_{-2/3}^{+}\rbrack \Pi _{0}+
\lbrack Q_{-2/3}^{+}Q_{-1/3}^{-}+Q_{-1}^{-}Q_{0}^{+}\rbrack \Pi
_{1}+\lbrack Q_{-2/3}^{-}Q_{-1/3}^{+}+Q_{-}^{+}Q_{0}^{-}\rbrack
\Pi _{-1}.
\end{equation}
Comparing Eqs (27) and (26), one finds the following constraint
relations:
\begin{equation}
\begin{array}{lcr}
Q_{-1/3}^{-}&=&Q_{-1/3}^{+}Q_{0}^{+}\nonumber \\
Q_{-1/3}^{+}&=&Q_{-1/3}^{-}Q_{0}^{-}\nonumber \\
Q_{-2/3}^{-}&=&Q_{0}^{+}Q_{-2/3}^{+} \\
Q_{-2/3}^{+}&=&Q_{0}^{-}Q_{-2/3}^{-}\nonumber \\
Q_{-1}^{-}&=&Q_{-2/3}^{+}Q_{-1/3}^{+}\nonumber \\
Q_{-1}^{+}&=&Q_{-2/3}^{-}Q_{-1/3}^{-}\nonumber. \\
\end{array}
\end{equation}
 Eqs (28) and consequently Eqs (26-27) may be satisfied identically by
introducing the following $3\times3$ matrix operator ${\bf Q}={\bf
Q}^{+}+{\bf Q}^{-}$ whose entries are
$Q_{-x}^{\pm}$'s,$x=0,1/3,2/3$ are the generators of FSS. This
matrix representation is in agreement with the ZF parafermionic
invariance property (22-23) and the constraint Eqs (28). Using the
gl(3) generators $E_{ij}=|i\rangle {\otimes} |j\rangle
$,$i,j=0,\pm 1$, one may write down the two following matrix
realisations for the $Q_{-x}^{\pm}$'s.\\ (i)$({1\over 3},0)$ Real
FSS: This algebra, to which we shall refer hereafter to  as
2d$({1\over 3},0)$ FSS, is the analogue of Eqs (2) and (5). The
matrix representation of the $Q_{-x}^{\pm}$'s reads as:
\begin{equation}
\begin{array}{lcr}
Q_{-2/3}^{+}&=&Q_{-2/3}E_{1,0}\nonumber \\
Q_{-2/3}^{-}&=&Q_{-2/3}E_{-1,0} \nonumber \\
Q_{0}^{+}&=&Q_{0}E_{-1,1}
\\ Q_{0}^{-}&=&Q_{0}E_{1,-1}\nonumber \\
Q_{-1/3}^{+}&=&Q_{-1/3}E_{0,-1}\nonumber \\
Q_{-1/3}^{-}&=&Q_{-1/3}E_{0,1}\nonumber. \\
\end{array}
\end{equation}
The charge carried by the  $Q_{-x}^{\pm}$ of Eqs (29) is the
$Z_{3}$ charge of the automorphism symmetry of the matrix operator
equation.
\begin{equation}
(E^{3})_{ii}=\Pi _{i};\quad i=0,\pm 1.
\end{equation}
In the orthonormal basis $\{|i\rangle , i=i=0,\pm 1\}$, the matrix
representation of the $Q_{-x}^{\pm}$'s reads as:
\begin{equation}
{\bf Q}^{+}=\left[
\begin{array}{ccccc}
0 & Q_{-2/3} & 0 \\ 0 &0 & Q_{0} \\ Q_{-1/3} &0 & 0 \\
\end{array}
\right] ;\quad \quad {\bf Q}^{-}=\left[
\begin{array}{ccccc}
0 & 0 & Q_{-2/3}   \\Q_{-1/3} & 0 & 0 \\0 & Q_{-1} & 0  ,
\end{array}
\right]  \label{matrices}
\end{equation}
Using Eqs (31) it is not difficult to check that the following
relation hold:
\begin{equation}
\begin{array}{lcr}
2{\bf P}={\bf Q}^{+}{\bf Q}^{-}+{\bf Q}^{-}{\bf Q}^{+}\quad \quad
(a)\nonumber
\\ {\bf Q}^{-}={\bf Q}^{+}{\bf Q}^{+}\quad \quad \quad \quad \quad \quad(b)\\
 {\bf Q}^{-2}={\bf P}.{\bf
Q}^{+}.\quad \quad \quad \quad \quad \quad(c)
\end{array}
\end{equation}
Taking the traces of both sides of these  matrix Eqs, one
discovers the algebra Eqs (17) which reads in terms of $Q_{-1/3}$
and $Q_{-2/3}$ as:
\begin{equation}
\begin{array}{lcr}
P_{-1}=Q_{-1/3}Q_{-2/3}+Q_{-2/3}Q_{-1/3}\nonumber \\
0=\{Q_{-1/3},Q_{-1/3}\}=\{Q_{-2/3},Q_{-2/3}\}.
\end{array}
\end{equation}
Note that Eq(33) was expected from the constraint Eq (7). It was
suggested in ref [ 8] as a linearized form of the non linear
operator equation $[Q_{-1/3}]^{3}=P_{-1}$. The relation between Eq
(33) and 2d$(({1\over 2}^{2}),0)$ supersymmetry suspected in [8]
will be considered latter on.\\ (ii)The $(({1\over 3}^{2}),0)$
FSA: This is a complex solution for which the matrix
representation of the  $Q_{-x}^{\pm}$'s reads as.
\begin{equation}
{\bf Q}^{+}=\left[
\begin{array}{ccccc}
0 & Q_{-2/3}^{+} & 0 \\ 0 &0 & Q_{0}^{+} \\ Q_{-1/3}^{+} &0 & 0 \\
\end{array}
\right] ;\quad \quad {\bf Q}^{-}=\left[
\begin{array}{ccccc}
0 & 0 & Q_{-2/3}^{-}   \\Q_{-1/3}^{-} & 0 & 0 \\0 & Q_{-1}^{-}& 0.
\end{array}
\right]  \label{matrices}
\end{equation}
Here also the charges carried by the $Q_{-x}^{\pm}$ are $Z_{3}$
charges. Similar calculations as for the $({1\over 3},0)$  algebra
show that Eqs (32) are again fulfiled for the representation (34).
Using Eqs (34) and solving Eqs (32), we find the following
relations :
\begin{equation}
\begin{array}{lcr}
2P_{-1}=\{Q_{-2/3}^{+},Q_{-1/3}^{-}\}+\{Q_{-1/3}^{+},Q_{-2/3}^{-}\}\quad
\quad (a)\nonumber
\\0=\{Q_{-1/3}^{\pm},Q_{-1/3}^{\pm}\}=\{Q_{-2/3}^{\pm},Q_{-2/3}^{\pm}\}\quad\quad\quad\quad(b)\nonumber
\\Q_{-1/3}^{-}=Q_{-1/3}^{+}Q_{0}^{+};\quad
Q_{-1/3}^{+}=Q_{0}^{-}Q_{-1/3}^{-}\quad \quad \quad (c) \\
Q_{-2/3}^{-}=Q_{0}^{+}Q_{-2/3}^{+}\nonumber;\quad
Q_{-2/3}^{+}=Q_{-2/3}^{-}Q_{0}^{-}\quad \quad \quad(d) .
\end{array}
\end{equation}
Eqs (35) define the $(({1\over 3})^{2},0)$ fractional
supersymmetrc algebra . It is generated by two $Z_{3}$  doublets
of anticommuting charges operators $Q_{-1/3}^{\pm}$ and
$Q_{-2/3}^{\pm}$. The components of each doublet are related to
each another by the $Q_{0}^{\pm}$'s as shown on Eqs (35-c-d). Note
that the algebra (35) is stable under the three following
conjugations:(a)$(Q_{-x}^{\pm})^{*}=Q_{-x}^{\mp}$;
(b)$(Q_{-x}^{\pm})^{{\bf +}}=Q_{-1+x}^{\mp}$ and
(c)$(Q_{-x}^{\pm})^{{\bf +}*}=Q_{-1+x}^{\pm}$ suggesting a link
with 2d N=$(({1\over 2})^{4},0)$su(2) supersymmetry formulated in
harmonic superspace [9]. In what follows we shall explore this
relation in order to use it for the construction of off shell
representations of the 2d$(({1\over 3})^{2},0)$ FSS.
 \section{More on the algebra equations (35)}

Because of the periodicity $(p\equiv p\pm 6)$   of the
representations of the  ZF parafermionic algebra (19) which allow
to identify the field operators $\Phi ^{\pm}$  as $\Phi
^{\pm}\equiv {\Phi ^{\pm}}^{2}$ and $\Phi ^{\pm \pm }\equiv {\Phi
^{\pm}}^{4}$ ; see Eq (20-b) one may rewrite the charge operators
$Q_{0}^{+}, Q_{0}^{-}$ as $Q_{0}^{--}, Q_{0}^{++}$, respectively.
Moreover denoting by $Q_{0}^{0}$ the commutator of the $Q_{0}^{+}$
and $Q_{0}^{-}$ charge operators, one sees from Eqs (29) that
$Q_{0}^{0}$ and $Q_{0}^{\pm}$ generate altogether an su(2)
algebra.
\begin{equation}
\begin{array}{lcr}
\lbrack Q_{0}^{-},Q_{0}^{+}\rbrack =Q_{0}^{0} ,\nonumber \\
\lbrack Q_{0}^{0},Q_{0}^{\pm}\rbrack =\mp Q_{0}^{\pm},
\end{array}
\end{equation}
acting on the $Q_{-x}^{\pm},x={1\over 3},{1\over 3}$  charges as:
\begin{equation}
\begin{array}{lcr}
\lbrack Q_{0}^{0},Q_{-1/3}^{\pm}\rbrack =\pm Q_{-1/3}^{\pm}
,\nonumber \\ \lbrack Q_{0}^{0},Q_{-2/3}^{\pm}\rbrack =\pm
Q_{-2/3}^{\pm} ,\nonumber \\ \lbrack
Q_{0}^{\pm},Q_{-1/3}^{\pm}\rbrack =\mp Q_{-1/3}^{\mp} ,
\\ \lbrack
Q_{0}^{\pm},Q_{-2/3}^{\pm}\rbrack =\pm Q_{-2/3}^{\mp}\nonumber \\
\lbrack Q_{0}^{-},Q_{-x}^{+}\rbrack =0 ,\nonumber
\\\lbrack Q_{0}^{+},Q_{-x}^{-}\rbrack =0 ,\nonumber .
\end{array}
\end{equation}
Eqs (36) correspond just to the zero mode subalgebra of the level
3 of the $su_{3}(2)$ Kac Moody symmetry. The latter  is known to
be homomorphic to the $Z_{3}$ parafermionic invariance (19) [10].
Now using  the identification  $Q_{0}^{\mp}\equiv Q_{0}^{\pm \pm
}$ and the $Z_{3}$ periodicity $q\equiv q mod(3)$, one may rewrite
the algebra (36) as
\begin{equation}
\begin{array}{lcr}
\lbrack Q_{0}^{++},Q_{0}^{--}\rbrack =Q_{0}^{0} ,\quad
(a)\nonumber
\\ \lbrack Q_{0}^{0},Q_{0}^{\pm \pm}\rbrack =\pm Q_{0}^{0}.\quad
 (b)
\end{array}
\end{equation}
Substituting $Q_{0}^{\mp}$ by $Q_{0}^{\pm \pm}$  in Eqs (37); one
gets the following relations which looks like the coresponding
ones in 2d N=4 su(2) supersymmetry [9]:
\begin{equation}
\begin{array}{lcr}
\lbrack Q_{0}^{++},Q_{-x}^{+}\rbrack =\lbrack
Q_{0}^{--},Q_{-x}^{-}\rbrack =0,\nonumber
\\  \lbrack Q_{0}^{++},Q_{-1/3}^{-}\rbrack =aQ_{-1/3}^{+}\nonumber
\\\lbrack Q_{0}^{--},Q_{-2/3}^{+}\rbrack =aQ_{-2/3}^{-}\nonumber
\\\lbrack Q_{0}^{0},Q_{-1/3}^{\pm}\rbrack =\pm aQ_{-1/3}^{\pm}
\\\lbrack Q_{0}^{--},Q_{-1/3}^{+}\rbrack =bQ_{-1/3}^{-}\nonumber
\\\lbrack Q_{0}^{++},Q_{-2/3}^{-}\rbrack =bQ_{-2/3}^{+}\nonumber
\\ \lbrack Q_{0}^{0},Q_{-2/3}^{\pm}\rbrack =\pm aQ_{-2/3}^{\pm},
\end{array}
\end{equation}
where a=-b=1. Recall that in 2d$(({1\over 3})^{2},0)$ su(2)
supersymmetry the coefficients a and b are equal to one, a=b=1,and
the analogue of Eqs (35-a-b) read as:
\begin{equation}
\begin{array}{lcr}
2P_{-1}=\{Q_{-1/2}^{+},\bar{Q}_{-1/2}^{-}
\}-\{Q_{-1/2}^{-},\bar{Q}_{-1/2}^{+} \} \quad  (a)\nonumber
\\
0=\{Q_{-1/2}^{\pm},Q_{-1/2}^{\pm}\}=\{\bar{Q}_{-1/2}^{\pm},\bar{Q}_{-1/2}^{\pm}\}
\quad \quad \quad(b).
\end{array}
\end{equation}
Observe by the way that apart from the spin of the charge
operators, Eq (38-a) differs from Eq (35-a) by the presence of the
minus sign which is required by invariance under the su(2)
automorphism group of the 2d N=4 su(2) supersymmetric algebra.
Nevertheless the similarity between Eqs (35) and (37-38) allows us
to build an off shell superspace formulation of 2d$(({1\over
3})^{2},0)$ supersymmetry by mimiking the harmonic superspace
formalism [11].
\section{Superfield theory.}
We start first by describing the superfield theory of the
2d$({1\over 3},{1\over 3})$, fractional supersymmetry Eq (33)
especially the matter couplings of the on shell scalar
representation $(\varphi ,\Psi _{\pm 1/3},\Psi _{\pm 2/3})$. Using
the formal analogy between Eqs (33) and those of 2d N=2 U(1)
supersymmetry, namely
\begin{equation}
\begin{array}{lcr}
Q_{\mp 1/3}&\longrightarrow &Q_{\mp 1/2}^{+}\nonumber \\ \theta
_{\pm 1/3}&\longrightarrow &\theta _{\pm 1/2}^{+}\nonumber \\
Q_{\mp 2/3}&\longrightarrow & Q_{\mp 1/2}^{-}\nonumber \\ \theta
_{\pm 2/3}&\longrightarrow & \theta _{\pm 1/2}^{-}\\ \varphi
&\longrightarrow &\varphi (y)\nonumber \\ \Psi _{\mp
1/3}&\longrightarrow &\Psi _{\mp 1/2}^{+}\nonumber \\ \varphi
&\longrightarrow &\varphi (y)\nonumber \\ \Psi _{\mp
2/3}&\longrightarrow &\Psi _{\mp 1/2}^{-},
\end{array}
\end{equation}
and following the same lines used in the   building of 2d N=2 U(1)
supersymmetric matter couplings [12], one sees that the superfield
action $S[\Phi ,\bar{\Phi}]$ invariant under the 2d$({1\over
3},{1\over 3})$ fractional supersymmetry reads as:
\begin{equation}
S[\Phi ,\bar{\Phi}]=\int d^{2}z d\theta _{1/3}d\theta
_{-1/3}d\theta _{2/3}d\theta _{-2/3}K[\Phi ,\bar{\Phi}].
\end{equation}
In this Eq, K is the Kahler potential depending on the chiral
superfields $\Phi $ and $\bar{\Phi}$ given by :
\begin{equation}
\begin{array}{lcr}
\Phi &=&\varphi +\theta _{1/3}\Psi _{-1/3}+\theta _{-1/3}\Psi
_{1/3}+\theta _{1/3}\theta _{-1/3}F,\nonumber \\ \bar{\Phi}
&=&\bar{\varphi} +\theta _{2/3}\Psi _{-2/3}+\theta _{-2/3}\Psi
_{2/3}+\theta _{2/3}\theta _{-2/3}F.
\end{array}
\end{equation}
$\Phi $ and $\bar{\Phi}$ describe a complex scalar representation
of the algebra (33); each bosonic degree of  freedom  has two
partners of spin ${1\over 3}$ and ${2\over 3}$. Note that using Eq
(42) and the matter superfield representation Eq (43); one gets
the right two point free correlation functions; i.e:$<\Psi
_{-1/3}(z_{1})\Psi _{-2/3}(z_{2})>$ and $<\Psi _{-1/3}(z_{1})\Psi
_{-1/3}(z_{2})>=<\Psi _{-2/3}(z_{1})\Psi _{-2/3}(z_{2})>=0$.
Concerning the $(({1\over 3})^{2},0)$ fractional supersymmetric
algebra (35) generated by the four generators $Q_{-1/3}^{\pm}$ and
$Q_{-2/3}^{\pm}$, one may use here also the similarity with 2d N=4
extended supersymmetry. Thus extending Eqs (35) as:
\begin{equation}
\begin{array}{lcr}
P_{-1}=\{D_{-2/3}^{+},D_{-1/3}^{-} \}=-\{D_{-1/3}^{+},D_{-2/3}^{-}
\}\nonumber
\\
0=\{D_{-x}^{\pm},D_{-y}^{\pm} \},
\end{array}
\end{equation}
together with
\begin{equation}
\begin{array}{lcr}
\lbrack D^{++},D_{-x}^{+}\rbrack =\lbrack D^{++},D_{-x}^{-}\rbrack
=0;\quad x={1\over 3},{2\over 3}\nonumber
\\ \lbrack D^{++},D_{-x}^{-}\rbrack =D_{-x}^{+}\nonumber \\
 \lbrack
D^{++},D^{--}\rbrack =D^{0}\\ \lbrack D^{0},D_{-x}^{\pm}\rbrack
=\pm D_{-x}^{\pm},
\end{array}
\end{equation}
and using the Grassman variables $\theta _{1/3}^{\pm}$ and $\theta
_{2/3}^{\pm}$ of spin ${1\over 3}$ and ${2\over 3}$ respectively,
but still satisfying ${\theta _{x}^{\pm}}^{2}=0$ one can build a
superspace realisation of Eqs (44-45) and consequently a 2d
quantum superfield theory. A remarkable realisation of the algebra
(44-45), stable under the combined conjugation ${\bf +}*$  and
using the covariant superderivatives $D _{x}^{\pm}$ and $D
_{x}^{\pm \pm}$, $D _{0}^{0}$ instead of the $Q _{-x}^{\pm}$
generators is given by:
\begin{equation}
\begin{array}{lcr}
D _{-1/3}^{+}=-{\partial \over \partial \theta
_{-1/3}^{-}}\nonumber
\\ D _{-2/3}^{+}=-{\partial \over \partial \theta
_{-2/3}^{-}}\nonumber \\
 D _{-2/3}^{-}={\partial  \over \partial
\theta _{2/3}^{+}}- \theta _{1/3}^{-} P_{-1}\nonumber \\ D
_{-1/3}^{-}={\partial  \over \partial \theta _{1/3}^{+}}- \theta
_{2/3}^{-} P_{-1}\\P_{-1}={\partial \over \partial y}\\
\end{array}
\end{equation}
where $y=z-{1\over 2}( \theta _{1/3}^{-}\theta _{2/3}^{+}+ \theta
_{1/3}^{+}\theta _{2/3}^{-})$ and where
\begin{equation}
\begin{array}{lcr}
D^{++}=\lbrack u^{+i}{\partial \over \partial u^{-i}}-\theta
_{1/3}^{+}\theta _{2/3}^{+}P_{-1}\rbrack \nonumber\\
D^{--}=\lbrack u^{-i}{\partial \over \partial u^{+i}}-\theta
_{1/3}^{-}\theta _{2/3}^{-}P_{-1}\rbrack \\ D^{0}=\lbrack
D^{++},D^{--}\rbrack .
\end{array}
\end{equation}
In this realisation $u_{i}^{\pm}$ are the well known harmonic
variable satisfying $u^{+i}u_{i}^{-}=1$ and $u^{\pm
i}u_{i}^{\pm}=0$. Using the realisation (46-47), the on shell
matter multiplet $(0^{4},({1\over 3})^{4},({2\over 3})^{4})$ of
the fractional supersymmetric algebra (44-45) is given by a
hermitean superfield satisfying the analytic conditions
\begin{equation}
D_{-1/3}^{+}\Phi =D_{-2/3}^{+}\Phi=0,
\end{equation}
and the equation of motion
\begin{equation}
{D^{++}}^{2}\Phi =0.
\end{equation}

The $\theta _{x}^{+}$ expanssion of the superfield $\Phi $, stable
under the combined conjugation $({\bf +}*)$ reads as:
\begin{equation}
\Phi =\varphi ^{ij}u_{i}^{+}u_{j}^{-}+\theta _{1/3}^{+}\Psi
_{-1/3}^{i}u_{i}^{-}+\theta _{2/3}^{+}\Psi
_{-2/3}^{i}u_{i}^{-}+\theta _{1/3}^{+}\theta
_{2/3}^{+}F^{ij}u_{i}^{-}u_{j}^{-}.
\end{equation}
The action of $S[\Phi ]$ describing the dynamics and the couplings
of the superfields $\Phi $ is similar to that of 2d$(({1\over
2})^{4},0)$  su(2) harmonic superspace. More generally the matter
couplings of the $(({1\over 3})^{2},({1\over 3})^{2}))$ su(2)
fractional supersymmetry, extending Eqs (44-45) by adjoining the
antianalytic part, read in the harmonic superspace formulation [ 9
] as:
\begin{equation}
S[\Phi ]=\int d^{2}zd^{2}\theta _{\pm 1/3}^{+}d^{2}\theta _{\mp
2/3}^{+}du L^{+4}(\Phi ),
\end{equation}
where $L^{+4}$ is a functional of the superfield $\Phi $.
\section{Discussions and conclusion}
Using conformal parafermionic field theoretical methods, we have
studied realisations of the fractional supersymmetric algebras
extending the usual 2d supersymmetries. We have found that all the
known difficulties present in models based on $U_{q}(sl(2))$ and
non commutative geometry representations are overpassed. One of
the consequences of the new approach is that instead of one
generator $Q_{-1/k}$, FSAs are generated by many basic charge
operators $Q_{-x},\quad x=1/k,1/2k,\ldots $,For K=3 for example,
the $({1\over 3},0)$FSA is generated by two main charge operators
$Q_{-1/3}$ and $Q_{-2/3}$ satisfying Eqs (33) and the $(({1\over
3})^{2},0)$ FSA is generated by two doublets $Q_{-1/3}^{\pm}$ and
$Q_{-2/3}^{\pm}$ verifying Eqs (35). The proliferation of the
fractional supersymmetric generators is the price one should pay
in order to build a 2d local fractional supersymmetric QFT. This
feature explains why the realisation obtained in ref [8] solves
the constraints of fractional supersymmetry. It tells us moreover
that there is a close connection between $({1\over 3},{1\over 3})$
and $(({1\over 3})^{2},({1\over 3})^{2}))$ FSA's and 2d N=2U(1)
and N=4 su(2) supersymmetric algebras respectively. On the other
hand the conformal field methods show us that fractional
supersymmetry has other remarkable features which become manifest
for $K\geq 3$. As an example, the standard definition of
fractional heterotic superalgebras as $Q^{K} = P;\quad k>2$ is a
formal one. The right way to define it is as $P = tr [Q^{K}]$
where Q is a $K\times K$ matrix. For K=3, we found that $Z_{3}$
invariant fractional supesymmetric algebra is generated by
$Q_{-x},\quad x={1\over 3},{2\over 3},\ldots $ together with $Q_{0
}^{\pm}$ rotating the components of these doublets. The energy
momentum vector operator $P_{-1}$ in the $Z_{3}$ FSA is just the
trace of the following $3\times 3$ matrix: $${\bf
Q}^{3}=Q_{0}^{+}Q_{-2/3}^{+}\Pi _{1}+Q_{-1/3}^{+}Q_{0}^{+}\Pi
_{2}+Q_{-2/3}^{+}Q_{-1/3}^{+}\Pi _{3},$$ where the $\Pi _{i}$ 's
are projectors on the states $|i\rangle,\quad i=1,2,3$. This new
result is different from the one used in the literature according
to which ${\bf Q}^{3}={\bf Q}_{-1/3}{\bf Q}_{-1/3}{\bf Q}_{-1/3}$
is a C-operator. Furthermore the method based on parafermions we
have been using in this study tells us also that fractional spin
objects may be treated as anticommuting objects. The point is that
all fractional spin objects involved in the game are valued in the
gl(3) nilpotent subalgebra and more precisely their square is
identically zero; see Eqs (29). This anticommuting feature is
reconfirmed by the braiding property of conformal fields which
shows that the field operators $\Psi _{\Delta _{1}}$ and $\Psi
_{\Delta _{2}}$ of fractional spin such that $\Delta _{1}+\Delta
_{2}=1$, should anticommute. In particular the correlation
function $\langle \Psi _{-1/3}(z_{1}),\Psi _{-2/3}(z_{2})\rangle$
is equal to $-\langle \Psi _{-2/3}(z_{1}),\Psi
_{-1/3}(z_{2})\rangle$;a fact which imply that one may build a
local 2d quantum field theory invarint under fractional
supersymmetry. In the end of discussion, we would like to note
that as far as locality is concerned, the building of a local 2d
fractional supersymmetric theory may be achieved by linearizing
the cubic relation ${\bf Q}^{3}=P_{-1}$. There are differents ways
to perform this linearisation. A way to do it is as we have done
in this study where Eqs (26) and (28) are remplaced by Eqs
(35-39). Another way is as done in [8]where the nilpotency
condition $\theta _{1/3}^{3}=0$ , $\theta _{1/3}^{2}\neq 0$,
necessary for the construction of generalized superspace formalism
is solved of two anticommuting Grassman variables $\theta
_{1/3}^{+}$ and $\theta _{2/3}^{-}$; $\theta _{1/3}^{+2}=\theta
_{2/3}^{-2}=0$. Though this linearisation works, no explanation
was given in [8]. In the present analysis this feature can be
derived in a rigourous way. the analogue of the higher order
nilpotency condition $\theta _{1/3}^{3}=0$, $\theta _{1/3}^{2}\neq
0$ is trivially satisfied here since $\theta _{1/3}^{+2}\theta
_{2/3}^{+}=\theta _{1/3}^{+}\theta _{1/3}^{+2}=0$.

\section*{Aknowledgements}
The authors would like to thank Dr A.Elfallah and Dr J.Zerouaoui
for stimulating discussions. This research work has been supported
by the programm PARS under contrat 372-98 CNR.


\newpage
\begin{thebibliography}{10}
\bibitem[1]{a} Leclair A and  Vafa C 1993 Nucl. Phys. B {\bf 401}
413\\ (Leclair A and  Vafa C C 1992 Preprint CLNS 92/1150, HUTP
 92/A045)\\
 Fateev V and  Zamolodchikov A B, Nucl. Phys. B {\bf 280}644\\
 Bais F,  Bouwknecht P,  Surridge M and Schoutens K 1988 Nucl. Phys. B {\bf
 304}348\\Bais F,  Bouwknecht P,  Surridge M and Schoutens K 1988 Nucl. Phys. B {\bf
 304}371\\
 Durand  S, 1993 Phys. Lett B {\bf 312} 115\\
 Durand  S, 1993 Mod. Phys. Lett A {\bf 8}2323
\bibitem[2]{b}  Bernard D and Leclair  A 1990 Nucl. Phys. B340 721\\
 C. Ahn, D. Bernard and A. Leclair, 1990 Nucl. Phys. B{\bf
       346}409\\
 Saidi E H,  Sedra M B and  Zerouaoui J, 1995 Class. Quantum
Grav. {\bf 12}1567.\\  Mussardo G 1992 Phys Rep. {\bf 218}  216
\bibitem[3]{c}    Perez A,  Rausch  de Traubenberg M and  Simon P 1996 Nucl.
Phys.B{\bf 482}325\\ (  Perez A,  Rausch  de Traubenberg M and
Simon P 1996 2D Fractional Supersymmetry for Rational Conformal
Field Theory: Application for Third-Integer Spin States,preprint
       LPT-96-07, hep-th/9603139)\\
Kadiri A,  Saidi E H, Sedra  M B and Zerouaoui J, 1994 On exotic
supersymmetries of the thermal deformation of minimal models
preprint IC/94/216
\bibitem[4]{d}    Elfallah A, Saidi  E H and  Zerouaoui J,1998 on finite dimentional
fractional superalgebras Preprint  UFR-HEP

\bibitem[5]{e}    Zamolodchicov A B and Fateev V A , 1985 Sov. Phys. JETP{\bf
62} 215\\
 Belavin A A, Polyakov A M, Zamolodchikov A B, 1984 Nucl. Phys. B{\bf 214} 333
\bibitem[6]{i}    Freidan D,  Qiu Z and  Shenker S, 1985 Phys. Lett. B{\bf 151}
37\\  Bershadsky M A, Knizhnik V G and  Teitelman M G, 1985 Phys.
Lett. B{\bf 151}31
\bibitem[7]{f}   Gepner D and  Qiu Z, 1987 Nucl. Phys. B{\bf 285}423\\
Chakir H,  Elfallah A and  Saidi E H, 1995 Mod. Phys. Lett {\bf
38}2931
\bibitem[8]{g}   Saidi   E H,  Sedra M B and Zerouaoui J , 1995 Class. Quantum.
Grav.{\bf 12} 2705

\bibitem[9]{h}  Sokatchev  E and  Stelle K S, 1987 Class. Quantum. Grav. {\bf
4}501\\
 Lhallabi T and  Saidi E H 1988 Int. J. Mod. Phys. A {\bf 3} 187

\bibitem[10]{k}   Elfallah A PhD Thesis LMPHE / 96
 Kakushadze Z and  Henry Tye S H, 1994 Phys Rev D{\bf 49} 4122

\bibitem[11]{l}    Galperin A Ivanov E, Ogievetsky V and Sokatchev E 1985 Class. Quantum Grav.{\bf 2} 617

\bibitem[12]{m}   Zumino B 1979 Phys. Lett. B{\bf 87} 203
\end{thebibliography}
\end{document}